\documentclass[english]{article}
\usepackage[latin9]{inputenc}
\usepackage{amsmath}
\usepackage{amssymb}
\usepackage{graphicx}

\makeatletter
\renewcommand{\textendash}{--}

\@ifundefined{showcaptionsetup}{}{%
 \PassOptionsToPackage{caption=false}{subfig}}
\usepackage{subfig}
\makeatother

\usepackage{babel}
\begin{document}

\title{A multi-sided generalization\\of the $C^{0}$ Coons patch}

\author{Péter Salvi\\
Budapest University of Technology and Economics}
\maketitle
\begin{abstract}
Most multi-sided transfinite surfaces require cross-derivatives at
the boundaries. Here we show a general $n$-sided patch that interpolates
all boundaries based on only positional information. The surface is
a weighted sum of $n$ Coons patches, using a parameterization based
on Wachspress coordinates.
\end{abstract}

\section{Introduction}

Filling an $n$-sided hole with a multi-sided surface is an important
problem in CAGD. Usually the patch should connect to the adjacent
surfaces with at least $G^{1}$ continuity, but in some applications
only positional ($C^{0}$) continuity is needed, and normal vectors
or cross derivatives at the boundary curves are not available.

For $n=4$, the $C^{0}$ Coons patch~\cite{Coons:1967} solves this
problem; in this paper we show how to generalize it to any number
of sides.

\section{Previous work}

Most transfinite surface representation in the literature assume $G^{1}$
constraints, and the patch equations make use of the fixed cross-derivatives
at the boundary. This can be circumvented by generating a \emph{normal
fence} automatically, e.g.~with a rotation minimizing frame~\cite{Wang:2008};
however, in a $C^{0}$ setting this is an overkill, simpler methods
exist.

One well-known solution is the harmonic surface, which creates a ``soap
film'' filling the boundary loop by solving the harmonic equation
on a mesh with fixed boundaries. This, however, minimizes the total
area of the surface, which often has unintuitive results, see an example
in Section~\ref{sec:Examples}.

The basic idea of the proposed method, i.e., to define the surface
as the weighted sum of $n$ Coons patches, each interpolating three
consecutive sides, is the same as in the CR patch~\cite{Salvi:2014}.

\section{The multi-sided $C^{0}$ Coons patch}

\begin{figure}
\begin{centering}
\includegraphics[width=0.4\columnwidth]{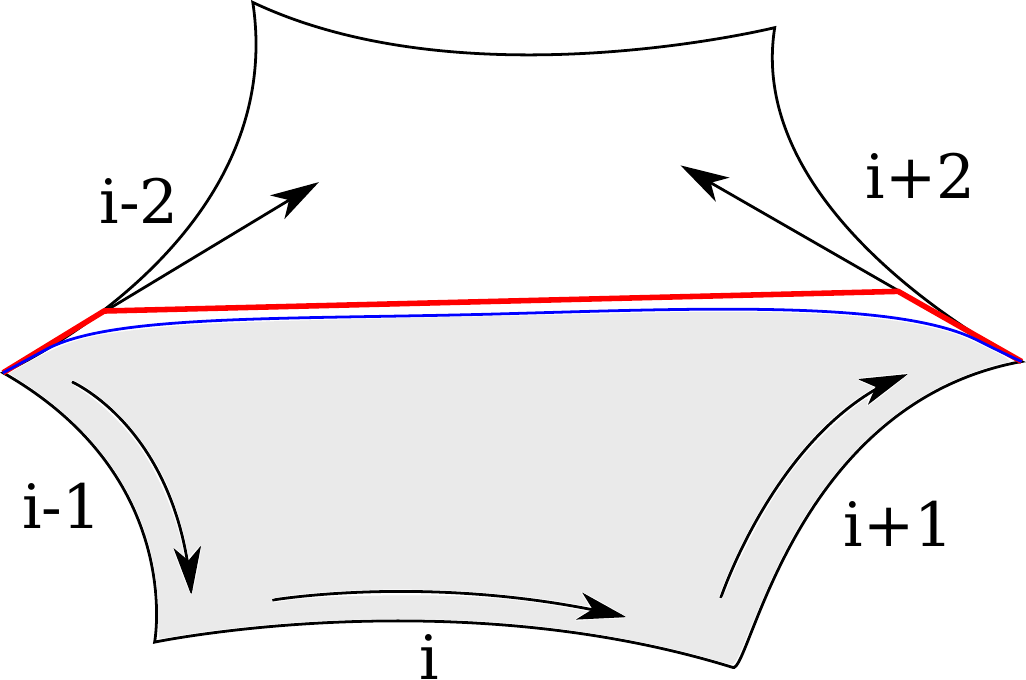}
\par\end{centering}
\caption{\label{fig:Construction-ribbon}Construction of a four-sided Coons
ribbon.}

\end{figure}

Let $C_{i}(t):[0,1]\to\mathbb{R}^{3}$ denote the $i$-th boundary
curve. Let us also assume $C_{i}(1)=C_{i+1}(0)$ for all $i$ (with
circular indexing). Then the \emph{ribbon} $R_{i}$ is defined as
a $C^{0}$ Coons patch interpolating $C_{i-1}$, $C_{i}$, $C_{i+1}$,
and $C_{i}^{\mathrm{opp}}$ \textendash{} a cubic curve fitted onto
the initial and (negated) end derivatives of sides $i+2$ and $i-2$,
respectively (see Figure~\ref{fig:Construction-ribbon}).
 
Formally,
\begin{align}
R_{i}(s_{i},d_{i}) & =(1-d_{i})C_{i}(s_{i})+d_{i}C_{i}^{\mathrm{opp}}(1-s_{i})\nonumber \\
 & +(1-s_{i})C_{i-1}(1-d_{i})+s_{i}C_{i+1}(d_{i})\nonumber \\
 & -\left[\begin{array}{c}
1-s_{i}\\
s_{i}
\end{array}\right]^{\intercal}\left[\begin{array}{cc}
C_{i}(0) & C_{i-1}(0)\\
C_{i}(1) & C_{i+1}(1)
\end{array}\right]\left[\begin{array}{c}
1-d_{i}\\
d_{i}
\end{array}\right],
\end{align}
where $C^{\mathrm{opp}}$ is defined as the Bézier curve\footnote{Except for $n=3$, where $C^{\mathrm{opp}}$ degenerates to the point
$C_{i+1}(1)$.} determined by the control points
\begin{align}
P_{0} & =C_{i+1}(1), & P_{1} & =P_{0}+\frac{1}{3}C_{i+2}'(0),\\
P_{2} & =P_{3}-\frac{1}{3}C_{i-2}'(1), & P_{3} & =C_{i-1}(0).
\end{align}

The surface is defined over a regular $n$-sided polygon. The Wachspress
coordinates of a domain point $p$ are defined as
\begin{equation}
\lambda_{i}=\lambda_{i}(p)=\frac{\prod_{j\neq i-1,i}D_{j}(p)}{\sum_{k=1}^{n}\prod_{j\neq k-1,k}D_{j}(p)},
\end{equation}
where $D_{i}(p)$ is the perpendicular distance of $p$ from the \mbox{$i$-th}
edge of the domain polygon. Ribbon parameterization is based on these
generalized barycentric coordinates:
\begin{align}
d_{i} & =d_{i}(u,v)=1-\lambda_{i-1}-\lambda_{i}, & s_{i} & =s_{i}(u,v)=\frac{\lambda_{i}}{\lambda_{i-1}+\lambda_{i}}.\label{eq:sd}
\end{align}
It is easy to see that $s_{i},d_{i}\in[0,1]$, and that $d_{i}$ has
the following properties:
\begin{enumerate}
\item $d_{i}=0$ on the $i$-th side.
\item $d_{i}=1$ on the ``far'' sides (all sides except $i-1$, $i$ and
$i+1$).
\item $d_{i-1}+d_{i+1}=1$ on the $i$-th side.
\end{enumerate}
Finally, we define the patch as
\begin{equation}
S(p)=\sum_{i=1}^{n}R_{i}(s_{i},d_{i})B_{i}(d_{i}),
\end{equation}
where $B_{i}$ is the blending function
\begin{equation}
B_{i}(d_{i})=\frac{1-d_{i}}{2}.
\end{equation}
The interpolation property is satisfied due to the properties of $d_{i}$
mentioned above.

(Note: $s_{i}$ in Eq.~(\ref{eq:sd}) cannot be evaluated when $d_{i}=1$,
but at these locations the weight $B_{i}(d_{i})$ also vanishes.)

\section{\label{sec:Examples}Examples}
\begin{figure*}
  {\hfill
\subfloat[Harmonic surface]{\begin{centering}
\includegraphics[width=0.4\textwidth]{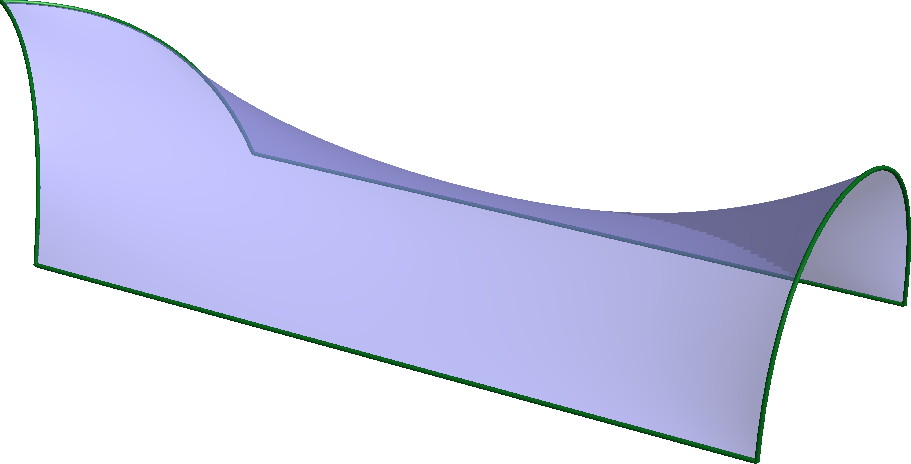}
\par\end{centering}
}\hfill{}\subfloat[$C^{0}$ Coons patch]{\begin{centering}
\includegraphics[width=0.4\textwidth]{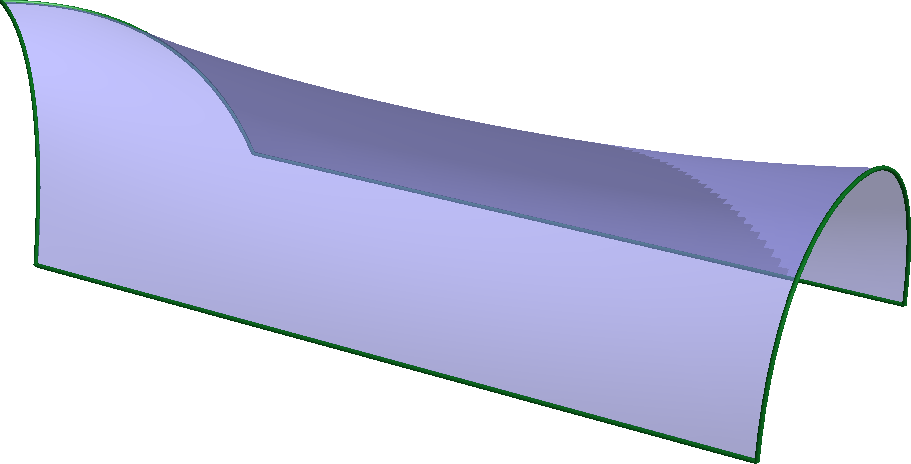}
\par\end{centering}
  }\hfill}
  \par
  {\hfill
\subfloat[Harmonic surface (contours)]{\begin{centering}
\includegraphics[width=0.4\textwidth]{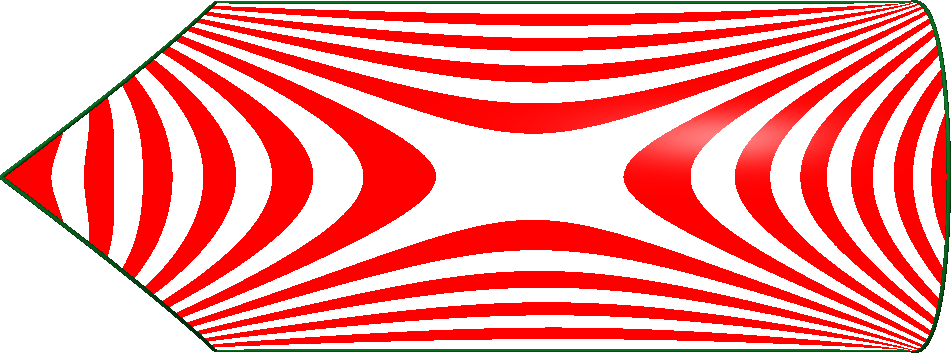}
\par\end{centering}
}\hfill{}\subfloat[$C^{0}$ Coons patch (contours)]{\begin{centering}
\includegraphics[width=0.4\textwidth]{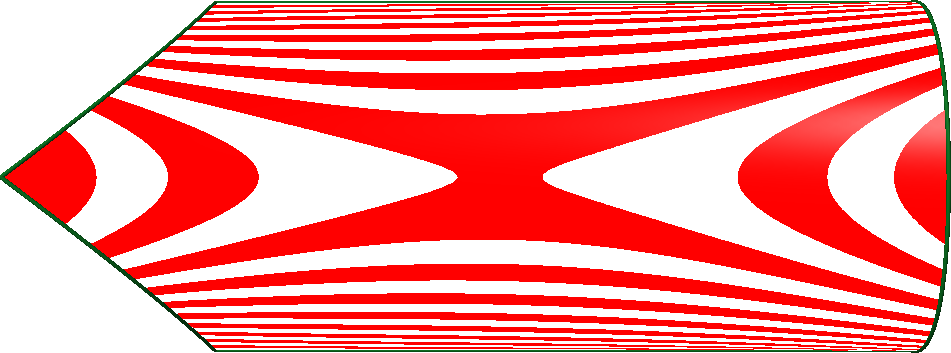}
\par\end{centering}
  }\hfill}

\caption{\label{fig:Comparison}Comparison with the harmonic surface on a 5-sided
boundary loop.}
\end{figure*}

Figure~\ref{fig:Comparison} shows a comparison with the harmonic
surface, which \textendash{} due to its area minimizing property \textendash{}
results in an unnaturally flat patch.

\begin{figure}
\subfloat[Mean curvature map]{\begin{centering}
\includegraphics[width=0.45\columnwidth]{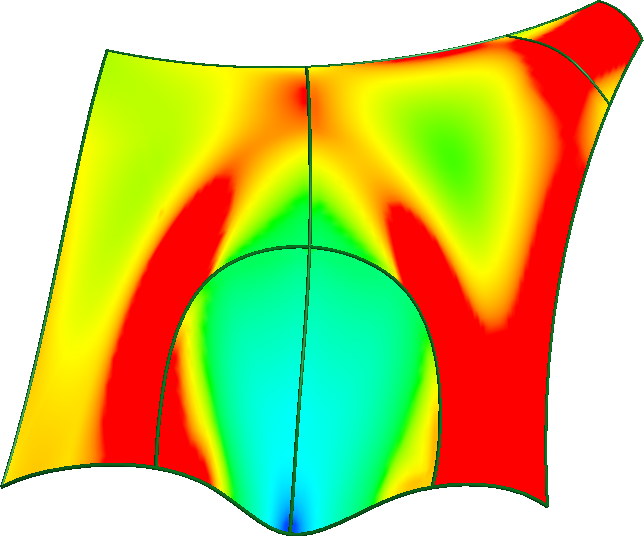}
\par\end{centering}
}
\hfill
\subfloat[Contouring]{\begin{centering}
\includegraphics[width=0.45\columnwidth]{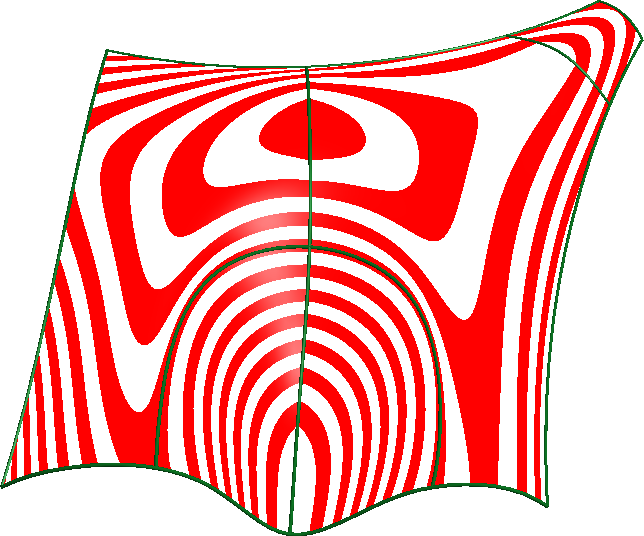}
\par\end{centering}}
\caption{\label{fig:pocket}The ``pocket'' model.}
\end{figure}

Figure~\ref{fig:pocket} shows a model with 5 patches: two 3-sided,
one 4-sided, one 5-sided, and one 6-sided. The mean curvature map
and contouring both show good surface quality.

\section*{Conclusion}

We have defined a natural generalization of the $C^{0}$ Coons patch
\textendash{} a lightweight and efficient multi-sided surface representation,
applicable when only positional data is available.

\section*{Acknowledgements}

This work was supported by the Hungarian Scientific Research Fund
(OTKA, No.\ 124727).\bibliographystyle{plain}
\bibliography{sajat,cikkek}

\end{document}